# Abundances and chemical stratification in the atmosphere of the HgMn star HD 175640

**M. Thiam[1,2], G. A. Wade[1], F. LeBlanc[2], V. R. Khalack[2]**

[1] *Department of Physics, Royal Military College of Canada, Kingston, Ontario, K7K 7B4, Canada*
[2] *Département de Physique et d'Astronomie, Université de Moncton, Moncton, N.-B., E1A 3E9, Canada*

**Abstract.**

We present the results of a study of the photospheric abundances of the HgMn star HD 175640, conducted using archival ESO-UVES spectra. A large number of unblended (titanium, chromium, manganese and iron) lines were studied to search for the presence of chemical stratification in the atmosphere of this star. The selected lines are located in the visible region of the spectrum, longward of the Balmer jump, in orders with S/N $\geq$ 300. We derived the abundance of each element by calculating independently the abundance associated each line. We then characterized the depth of formation of each line, and examined the dependence of abundance on optical depth. Titanium, chromium, manganese and iron show no variation of their abundance with optical depth. These four elements do not appear to be strongly stratified in the atmosphere of HD 175640. This indicates that if stratification occurs, it must be in atmospheric layers which are not diagnosed by the spectral lines studied, or that it is too weak to detect using these data. We also report evidence that HD 175640 is an SB1, and furthermore report anomalous shifts of some strong Fe II lines, the origin of which is unclear.

**Key words:** stars: abundances - stars: atmospheres - stars: chemically peculiar - stars: individual: HD 175640

## 1. Introduction

The mercury-manganese stars (HgMn) are of late B spectral type with effective temperatures ranging from 10500 K - 15000 K. These stars do not host detectable magnetic fields, with upper limits on longitudinal magnetic fields as small as about 5 G (Shorlin et al. 2002, Wade et al. 2006). They are characterized spectroscopically by extreme overabundances of mercury (possibly more than 6 dex; Cowley et al. 2006) and of manganese (possibly more than 3 dex), and more subtle over- and under-abundances of other elements. Considering the hydrodynamic stability of their atmospheres due to their slow rotation, these chemical anomalies are naturally interpreted as due to radiatively-driven diffusion and gravitational settling (e.g. Michaud 1970).

The aim of this paper is to report first results of a careful search for stratification in a small sample of sharp-lined HgMn stars. For one star in our sample, the B9 star HD 175640, we have derived the elemental abundances of titanium, chromium, manganese and iron, and derived the vertical distributions of these chemical elements in the stellar atmosphere.

## 2. Observations and spectrum reduction

A single spectrum of HD 175640, obtained on June 13, 2001 at ESO with the VLT UV-Visual Echelle Spectograph UVES at the UT2 telescope, was obtained from the ESO Science Archive. The spectrum covers the spectral domain from 3040 Å to 10000 Å with a resolving power ranging between 90 000 and 110 000. The data were reduced using the UVES pipeline data reduction software (version 2.2.0). Individual orders were then normalized using the IRAF "continuum" procedure. The signal-to-noise ratios of individual orders after normalization ranged between 200 to 600. For the high precision required for this investigation, we only considered the visible region of the spectrum, longward of the Balmer jump in orders with S/N $\geq$ 300. The stellar and laboratory wavelengths were compared to determine the stellar radial velocity. The latter was estimated to be (-34.2 $\pm$0.3) km s$^{-1}$. The spectrum of this star was compared to another spectrum obtained by J.D. Landstreet using the Gecko spectrograph at CFHT. Although the line profiles appear to have constant shape, clear ($\sim$ 10 km/s), systematic radial velocity differences are apparent. It thus appears that HD 175640 is a single-lined spectroscopic binary (SB1).

## 3. Spectrum synthesis

The atmospheric parameters of HD 175640 have been determined in the past by Hubrig et al. (1999). We have generated synthetic line profiles using the Zeeman2 code (Wade et al. 2001) and an ATLAS9 solar abundance atmosphere model corresponding to their results (12000g40). Comparing the computed profiles with the observed lines and calculating the reduced $\chi^2$, we observed that some lines were poorly fitted using this atmospheric model. By increasing the temperature of the atmosphere model by 500 K, we obtained $\chi^2$ values which are are closer to unity for the 12500g40 atmospheric model than for hotter or cooler models (by a factor larger than 1.5 times). The abundances derived with the 12000 K and 13000 K models are larger (by more than 0.2 dex) than the abundances obtained for 12500 K.

Ultimately, the stellar parameters adopted for the analysis of HD 175640 were effective temperature $T_{\rm eff}$ = 12500 K, surface gravity $\log g$ = 4.0, rotational velocity $v\sin i$ = 2.5 km s$^{-1}$ and a microturbulent velocity $\xi$ = 0.0 km s$^{-1}$. The modeling was carried out assuming an ATLAS9 model correponding to these parameters.

To characterize the depth of formation of each line, we assumed that each line formed at optical depth log $\tau_l$ = 0 at line center. The details of these calculations are described by Khalack et al. (2006).

Here we present the results of the analysis of titanium, chromium, manganese and iron lines. For each element, abundances were inferred from each individual line assuming a (vertically uniform) abundance in the atmosphere of HD 175640. Lines were carefully pre-selected to avoid blends and misidentifications. The atomic data for the line profile synthesis were extracted from the Vienna Atomic Line Database (VALD).

## 4. Results

A large number of lines (Table 1) was studied to search for stratification in the atmosphere of HD 175640. Panels a) and b) of Figure 1 show observed and computed profiles of iron lines with good-quality fits. However, some lines in the spectrum present anomalous radial velocity shifts (after correction for the systematic radial velocity); these are illustrated in panels c) and d) of Figure 1. The observed lines in panel c) present smaller shifts, whereas those in panel d) display larger shifts. The source of these shifts is currently unclear.

In Figure 2, for iron lines, we show the radial velocity difference between the observed and the adopted $V_r$ (of -34.2 km s$^{-1}$). For most of the lines, the difference ranges between -0.7 to 0.7 km

Table 1: Abundances $\log(N_{elem}/N_{tot})$ with standard deviations $\sigma$ and the number $n$ of lines measured for HD 175640.

| Ion | measured | $\sigma$ | n | Sun |
|---|---|---|---|---|
| Ti II | -5.71 | 0.17 | 29 | -7.14 |
| Cr I | -5.19 | 0.24 | 2 | -6.40 |
| Cr II | -5.51 | 0.24 | 42 | -6.40 |
| Mn I | -4.17 | 0.21 | 17 | -6.65 |
| Mn II | -4.41 | 0.21 | 43 | -6.65 |
| Fe I | -4.69 | 0.12 | 6 | -4.59 |
| Fe II | -4.84 | 0.12 | 60 | -4.59 |

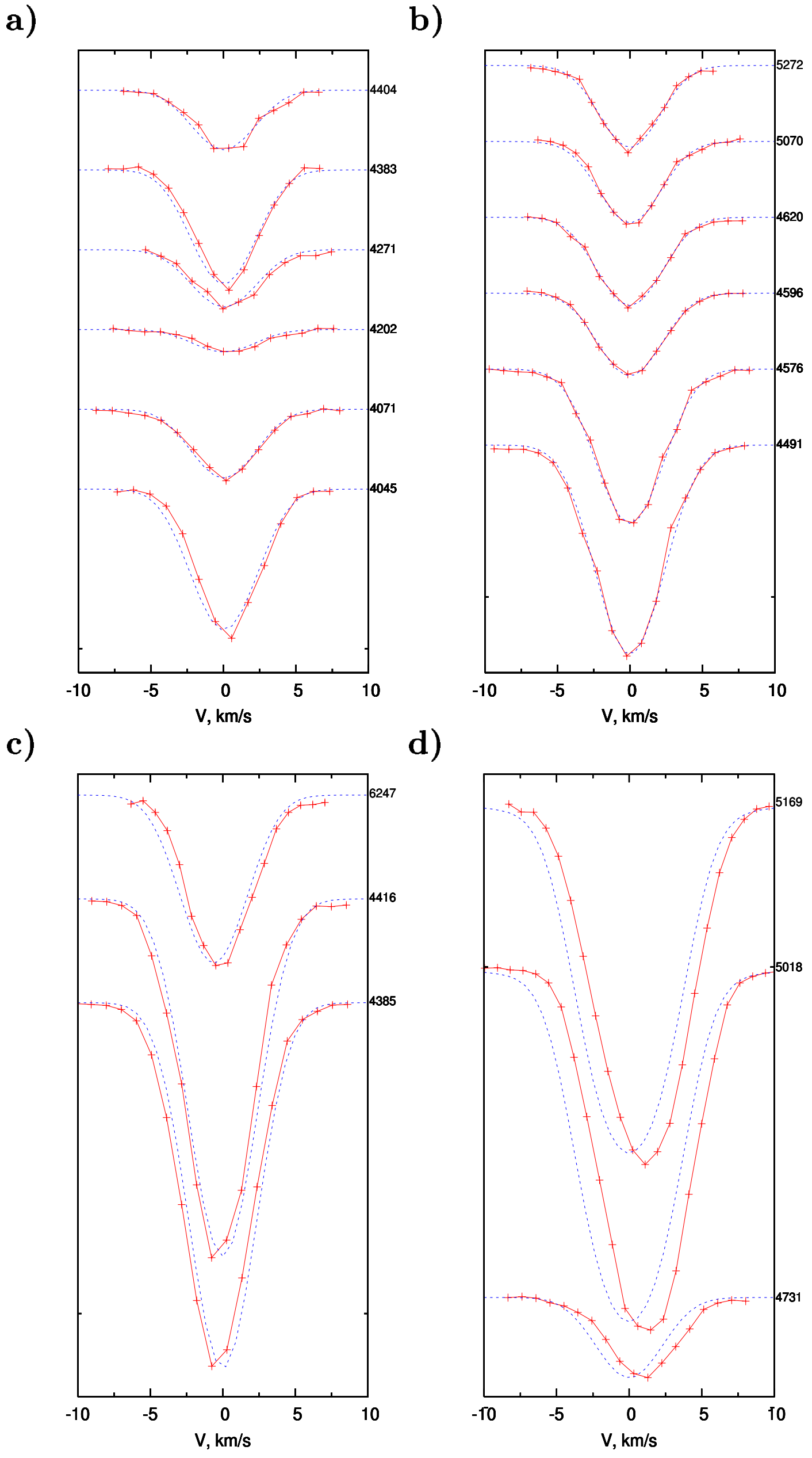

Figure 1: A comparison between the observed line profiles (full line) and computed lines (dashed line) of : a) neutral iron lines - b), c) and d) singly-ionized iron lines.

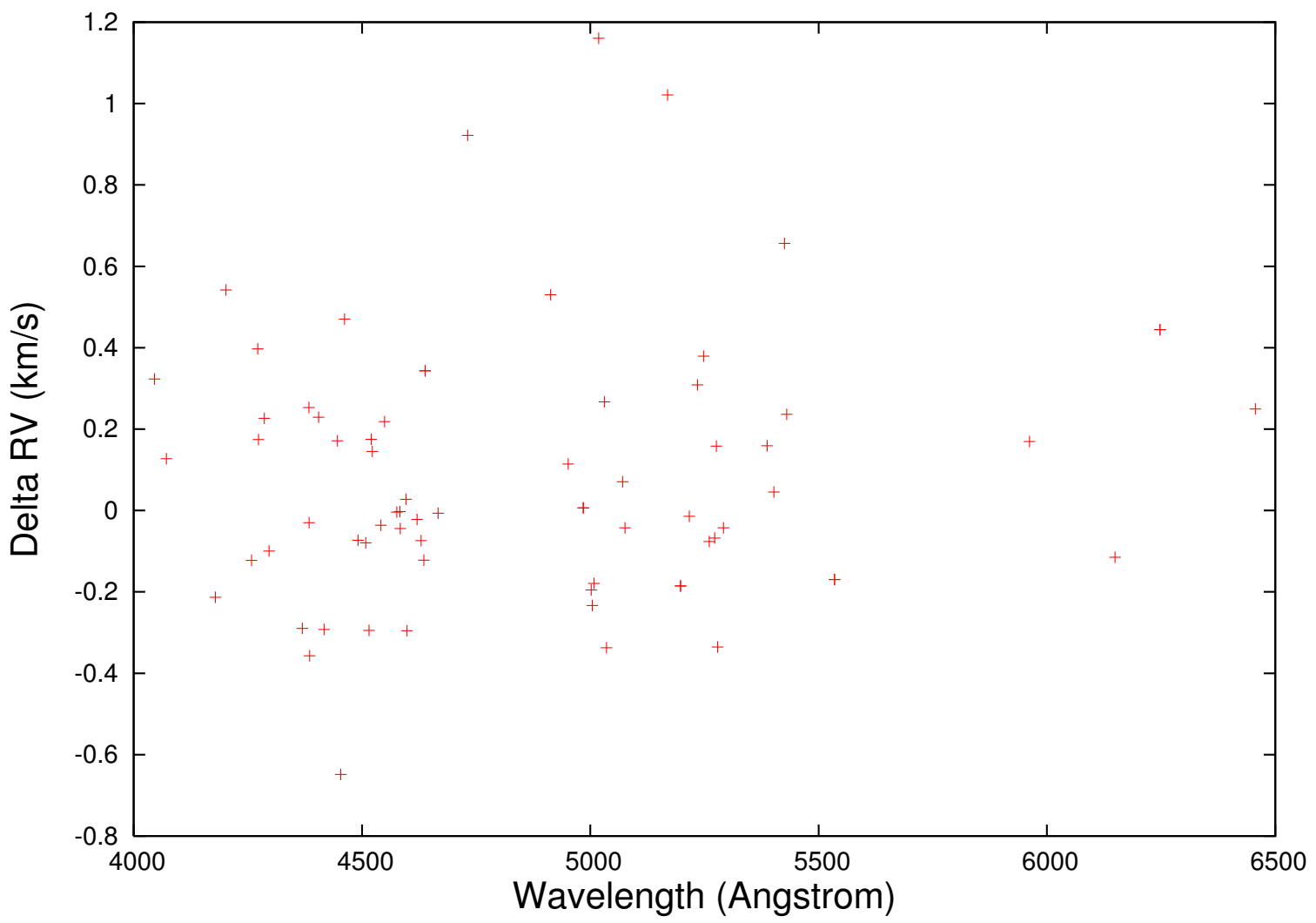

Figure 2: The radial velocity difference between the observed and the adopted radial velocities of iron lines.

s$^{-1}$. However, as shown in panel d) of Figure 1, the Fe II $\lambda\lambda\lambda$ 4731.453, 5018.440, 5169.033 lines present shifts larger than 0.8 km s$^{-1}$. A careful examination of the data reduction procedure and comparison with other spectra of HD 175640 indicates that these shifts are not instrumental in origin, nor do they results from the data reduction procedure. We are currently investigating if this reflects uncertainties (of $\sim$0.5-1 km s$^{-1}$) in published wavelengths of these lines, or if another (astrophysical) effect is responsible (e.g. binarity).

The remainder of this section reports results for each of the four elements studied.

*Titanium:* In Figure 3 (panel a), the titanium abundance decreases marginally with (lover level) excitation potential between 1-5.5 eV. However, this decrease is not large. In panel a) of Figure 4, the titanium abundance does not show any systematic variation in the optical depth range log $\tau_{5000}$ = -6 to -3. The mean abundance obtained is $-5.71 \pm 0.17$ dex. Titanium does not appear to be stratified in the optical depth range log $\tau_{5000}$ = -6 to -3 in the atmosphere of HD 175640.

*Iron:* The iron abundance shows no systematic variation as a function of (lower level) excitation potential in panel b) of Figure 3. Lookingat panel d) of Figure 4, we do not observe any significant variation of the abundance of Fe as function of $\tau_{5000}$ optical depth. We derived mean abundances of Fe I and Fe II, respectively, of -4.69 and -4.84 dex and a standard deviation of 0.12 dex. According to these results, iron shows no evidence of stratification in the atmosphere of HD 175640 in the optical depth range log $\tau_{5000}$ = -5.5 to -0.5.

*Chromium:* The chromium abundance, as illustrated in Figure 4 (panel b), is constant in the logarithmic optical depth range between -6 to 1. The mean abundances of Cr I and Cr II are respectively $-5.19 \pm 0.24$ and $-5.51 \pm 0.24$ dex. There is no sign of chromium abundance stratification in the atmosphere of this star.

*Manganese:* In panel c) of Figure 4, we show abundances derived from Mn lines. To model lines of Mn, hyperfine splitting was taken into account when data was available. Otherwise, observed lines were compared with computed lines neglecting hyperfine structure, and those lines showing relatively large reduced $\chi^2$ were not considered in the analysis. We obtained mean abundances of -4.17 and -4.41 dex, respectively for Mn I and Mn II and a standard deviation of 0.21 dex. We do not observe any systematic variation in manganese abundance in the optical depth range log $\tau_{5000}$ from -4.5 to -0.5.

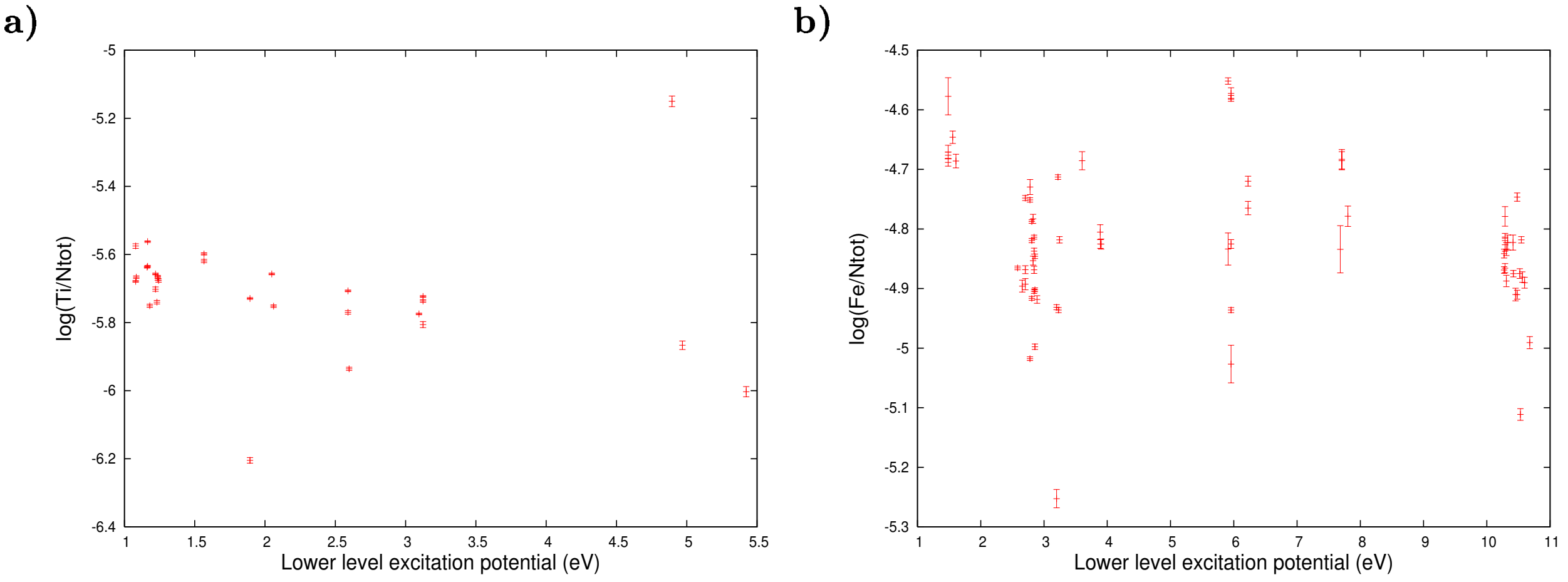

Figure 3: Dependence of the abundance on the lower level excitation potential for: a) titanium and b) iron.

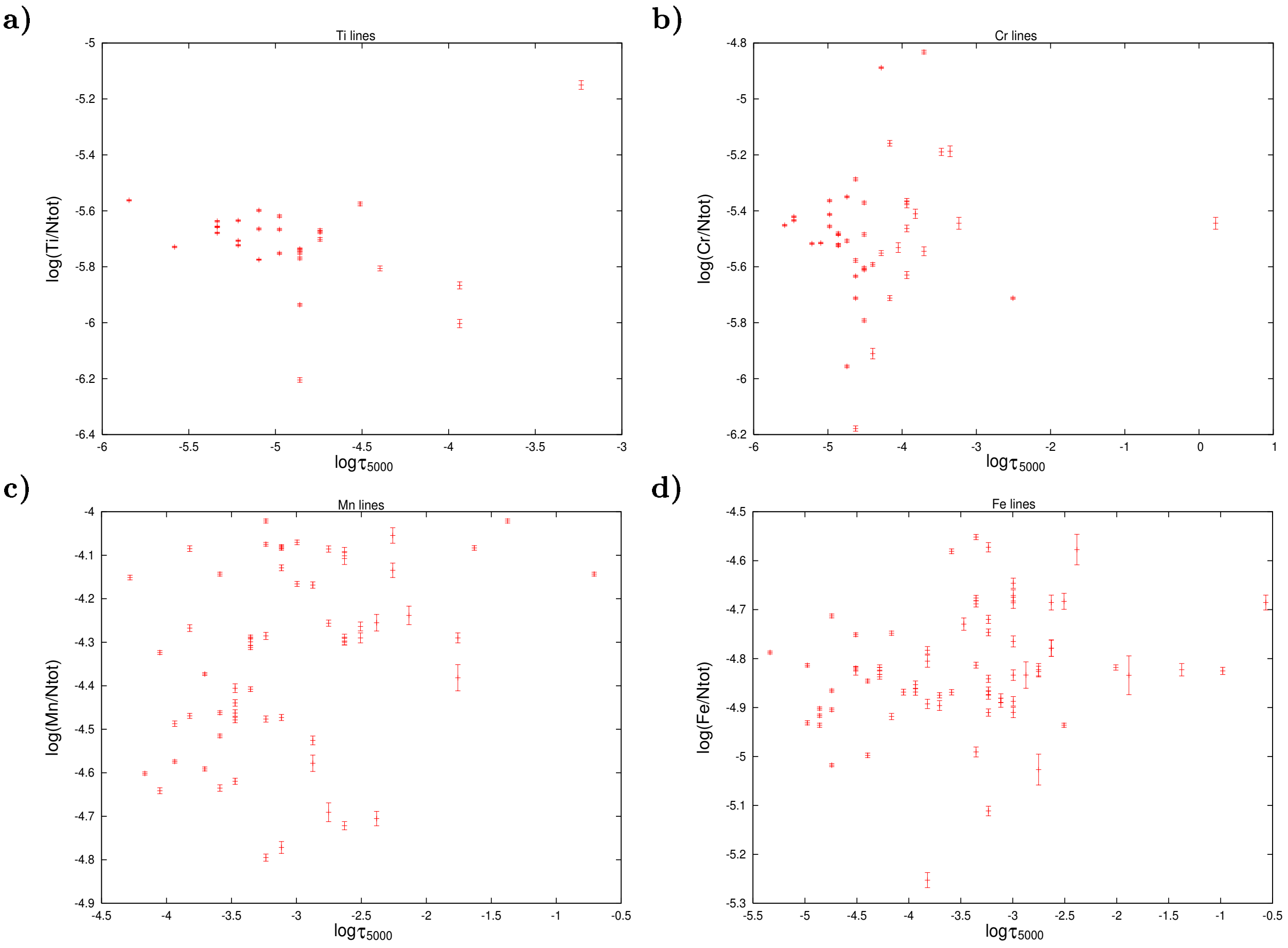

Figure 4: Dependence of the abundance on the optical depth for: a) titanium, b) chromium, c) manganese and d) iron.

## 5. Discussion

In this paper we have presented first results from a search for evidence of chemical stratification in the atmosphere of HD 175640. As presented in Table 1, the standard deviation in abundance for each element is less than 0.24 dex, which is at the level expected for uncertainties in $gf$-values and other atomic data. We have found no evidence for the stratification of titanium, chromium, manganese and iron in the atmosphere of this star, respectively in the optical depth ranges in log $\tau_{5000}$ from -6 to -3, -6 to 1, -4.5 to -0.5 and -5.5 to -0.5. For the All of the measured abundances in this study are consistent with those published by Castelli & Hubrig (2004).

We have observed shifts between observed and computed profiles of a number of lines for each elements which are not instrumental, and that we are unable to explain. Could these shifts be attributable to errors in the VALD database? On the other hand, could the effect be astrophysical, and consequently, might similar shifts be detectable in all sharp-lined HgMn stars?

Four other late-B stars (HD 71066, HD 90264, HD 178065 and HD 221507) are selected for the future investigations of stratification in the atmospheres of HgMn stars. Titanium, chromium, manganese, iron and yttrium will be studied for a search of stratification in these stars.

Finally, the measured abundances and line profiles will be compared with the predictions of self-consistent diffusion model atmospheres (e.g. Hui-Bon-Hoa et al. 2000).

## References

Castelli F. & Hubrig S., 2004, A&A, 425, 263
Hui-bon-Hoa A., LeBlanc F., Hauschildt P., 2000, ApJ 535, 43
Khalack V. R., LeBlanc F., Bohlender D., Wade G. A., Behr B. B., 2006, A&A
Michaud G., 1970, ApJ 160, 641
Shorlin S., Wade G.A., Donati J.-F., Landstreet J.D. et al., 2002, A&A 392, 637
Wade G, A, Bagnulo S., Kochukov O., et al. 2001, A&A, 375, 265
Wade G.A., Auriere M., Bagnulo S., Donati J.-F., et al., 2006, A&A 451, 293